\documentclass[12pt]{article}
\usepackage{epsfig,amssymb,times}
\newcommand{\ga}{\alpha}
\newcommand{\gb}{\beta}
\newcommand{\gc}{\gamma}

\newcommand{\gl}{\lambda}

\newcommand{\gL}{\Lambda}

\newcommand{\vk}[1]{\overline{#1}}
\newcommand{\uk}[1]{\underline{#1}}

\begin{document}
\title{Light front field theory\\ of quark matter at finite
  temperature\footnote{Based on invited talks presented at the Workshop on ``Effective
    quark quark interactions'' in Bled (Slovenia) July 7-14, 2003, and the
    310th
 WE-Heraeus Seminar ``Quarks in Hadrons and Nuclei II'', Rothenfels
    Castle, Oberw\"olz (Austria) September 15-20, 2003.}}
\author{Michael Beyer\\
  Fachbereich Physik, University of Rostock, D-18051 Rostock, Germany}
\date{}
\maketitle
\begin{abstract}
  A light front field theory for finite temperature and density is currently
  being developed. It will be used here to describe the transition region from
  quark matter to nuclear matter relevant in heavy ion collisions and in the
  early universe. The energy regime addressed is extremely challenging, both
  theoretically and experimentally. This is because of the confinement of
  quarks, the appearance of bound states and correlations, special relativity,
  and nonlinear phenomena that lead to a change of the vacuum structure of
  quantum chromodynamics. In the region of the phase transition it eventually
  leads to a change of the relevant degrees of freedom. We aim at describing
  this transition from quarks to hadronic degrees of freedom in a unified
  microscopic approach.
\end{abstract}

\section{Introduction}

Lattice calculations of quantum chromodynamics (QCD) give firm evidence that
nuclear matter undergoes a phase transition to a plasma state at a certain
temperature $T_c$ of about 170 to 180 MeV. Calculations have been performed at
a chemical potential $\mu=0$. Recent results are for staggered
fermions~\cite{Bernard:2003dk,Karsch:2000kv} and renormalization group
improved Wilson fermions~\cite{AliKhan:2001ek}.  The low density region
reflects, e.g., the scenario during the evolution of the early universe.  To
achieve information from lattice calculations at small $\mu$ several methods
have recently been developed, i.e., multiparameter
reweighting~\cite{Fodor:2001au,Fodor:2002km}, Taylor expansion at $\mu\simeq
0$~\cite{Allton:2002zi,Allton:2003vx}, imaginary
$\mu$~\cite{deForcrand:2002ci,D'Elia:2002gd,deForcrand:2003hx}.  The region of
validity is approximately $\mu\lesssim T$~\cite{Katz:2003up}. Effective
approaches to QCD indicate an extremely rich phase diagram also for $\mu>
T$~\cite{Alford:2002ng}.  Experimentally the QCD phase diagram is accessible
to heavy ion collisions. In particular relativistic heavy ion collision at
SPS/CERN and RHIC/BNL explore the region where hadronic degrees of freedom are
expected to be dissolved. Some results of RHIC are now available that give
hints of a non-hadronic state of matter~\cite{Heinz:2002gs}.

On the other hand light-front quantization of QCD can provide a rigorous
alternative to lattice QCD~\cite{Brodsky:1997de}.  Although the calculational
challenge in real QCD of 3+1 dimension seems large (as does lattice QCD) it
starts also from the fundamental QCD. The light-front quantization of QCD has
the particular advantage that it is completely formulated in physical degrees
of freedom. It has emerged as a promising method for solving problems in the
strong coupling regime.  Light front quantization makes it possible to
investigate quantum field theory in a Hamiltonian formulation~\cite{Dirac:cp}.
This makes it well suited for its application to systems of finite temperature
(and density). The relevant field theory has to be quantized on the light
front as well, which is presently being
developed~\cite{Beyer:2001bc,Mattiello:2001vq,Brodsky:2001wx,Brodsky:2001ww,Brodsky:2002,Alves:2002tx,Mattiello:2002ax,Weldon:2003uz,Weldon:2003vh,Das:2003mf,Kvinikhidze:2003wc,Beyer:2003qb}.
I present the light-front field theory at finite temperatures and densities in
the next section. For the time being it is applied to the Nambu-Jona-Lasinio
model (NJL) model~\cite{Nambu:tp,Nambu:fr} that is a powerful tool to
investigate the non-perturbative region of QCD as it exhibits spontaneous
breaking of chiral symmetry and the appearance of Goldstone bosons in a
transparent way. Finally, going a step further I shall give the general
in-medium light cone time ordered Green functions that allow us to treat quark
correlations that lead to hadronization.

 \section{Light front thermal field theory}

\label{sec:statphys}

  The four-momentum operator $P^\mu$ on the
light-front  is given by (notation of Ref.~\cite{Brodsky:1997de})
\begin{equation}
P^\mu=\int d\omega_+\; T^{+\mu}(x),
\label{eqn:P}
\end{equation}
where $T^{\mu\nu}(x)$ denotes the energy momentum tensor defined through the
Lagrangian of the system and $S_\mu$ is the quantization surface. The
Hamiltonian is given by $P^-$.  To investigate a grand canonical ensemble we
need the number operator,
\begin{equation}
N=\int d\omega_+\; j^+(x),
\label{eqn:N}
\end{equation}
where $j^\nu(x)$ is the conserved current. These are the necessary
ingredients to generalize the covariant partition operator at finite
temperature~\cite{Israel:1976tn,isr81,Weldon:aq} to the light-front.
The grand canonical partition operator on the
light-front is given by
\begin{equation}
Z_G = \exp\left\{\int d\omega_+\;
[- \gb_\nu T^{+\nu}(x)+\ga J^+(x) ]\right\},
\end{equation}
where $\ga=\mu/T$, with the Lorentz scalars temperature $T$ and chemical
potential $\mu$. The velocity of the medium is given by the time-like vector
$u_\nu u^\nu =1$~\cite{Israel:1976tn}, and $\gb_\nu = u_\nu/T$.  We choose the
medium to be at rest, $u^\nu = (u^-,u^+,\vec u^{\perp})= (1,1,0,0)$. The grand
partition operator then becomes
\begin{equation}
Z_G =  e^{-K/T},\quad K\equiv\frac{1}{2}(P^-+P^+) - \mu N
\label{eqn:ZG}
\end{equation}
with $P^\pm$ and $N$ defined in (\ref{eqn:P}) and (\ref{eqn:N}). 
The density operator for a grand canonical
ensemble~\cite{kad62,fet71}  in equilibrium
follows
\begin{equation}
\rho_G=(\mathrm{Tr}\ e^{-K/T})^{-1}\;e^{-K/T}.
\label{eqn:grand}
\end{equation}
The corresponding Fermi distribution
functions of particles $f^+\equiv f$ and antiparticles $f^-$ are given by
\begin{equation}
f^\pm(k^+,\vec k_\perp)=
\left[\exp\left\{\frac{1}{T}\left(\frac{1}{2}k^-_{\mathrm{on}}
+\frac{1}{2}k^+\mp\mu\right)\right\}+1\right]^{-1}
\label{eqn:fermipm}
\end{equation}
and $k^-_\mathrm{on}=(\vec k_\perp^2+m^2)/k^+$. This fermionic distribution
function (for particles) on the light-front has first been given
in~\cite{Beyer:2001bc}. The fermi function for the canonical ensemble can be
achieved by simply setting $\mu=0$. This then coincides with the distribution
function given recently in Ref.~\cite{Alves:2002tx} (up to different
metric conventions).

The  light-front time-ordered Green function for fermions is
\begin{equation}
i{\cal G}_{\ga\gb}(x-y)=
\theta(x^+-y^+)\;\langle\Psi_\ga(x)\bar\Psi_\gb(y)\rangle
- \theta(y^{+}-x^{+})\;\langle\bar\Psi_\gb(y)\Psi_\ga(x)\rangle.
\label{eqn:defCrono}
\end{equation}
We note here that the light-cone time-ordered Green function differs from the
Feynman propagator $S_F$ in the front form by a contact term $\gamma^+/2k^+$
and therefore coincides with the light-front propagator given previously in
Ref.~\cite{chang}.  To evaluate the ensemble average
$\langle\dots\rangle=\mathrm{Tr}(\rho_G\dots)$ of (\ref{eqn:defCrono}), we
utilize the imaginary time formalism~\cite{kad62,fet71}. We rotate the
light-front time of the Green function to imaginary value.  Hence the
$k^-$-integral is replaced by a sum of light-front Matsubara frequencies
$\omega_n$ according to~\cite{Beyer:2001bc},
\begin{eqnarray}
  \frac{1}{2} k^- \rightarrow i\omega_n - 
\frac{1}{2} k^+ +\mu\equiv \frac{1}{2} k_n^-\rightarrow\frac{1}{2}z,
\end{eqnarray}
where $\omega_n=\pi \lambda T$, $\lambda=2n+1$ for fermions [$\lambda=2n$ for
bosons]. In the last step we have performed an analytic continuation to the
complex plane.
For noninteracting Dirac fields
the (analytically continued) imaginary time Green function becomes
\begin{eqnarray}
G(z,\vk{k})&=&
\frac{\gamma k_\mathrm{on}+m}{z- k^-_\mathrm{on}+i\varepsilon}
 \frac{\theta(k^+)}{k^+}
(1-f^+(\vk{k}))
+\frac{\gamma k_\mathrm{on}+m}{z- k^-_\mathrm{on}-i\varepsilon} 
\frac{\theta(k^+)}{k^+}
f^+(\vk{k})
\label{eqn:lfmed}\\
&&
+\frac{\gamma k_\mathrm{on}+m}{z- k^-_\mathrm{on}+i\varepsilon} 
\frac{\theta(-k^+)}{k^+}
f^-(-\vk{k})
+\frac{\gamma k_\mathrm{on}+m}{z- k^-_\mathrm{on}-i\varepsilon} 
\frac{\theta(-k^+)}{k^+}
(1-f^-(-\vk{k})),
\nonumber
\end{eqnarray}
where $\vk{k}=(k^+,\vec k_\perp)$.  For equilibrium the imaginary time
formalism and the real time formalism are linked by the spectral
function~\cite{kad62,fet71,Kvinikhidze:2003wc}. For $\mu=0$ this propagator
coincides with that of~\cite{Kvinikhidze:2003wc}, but differs from that
of~\cite{Alves:2002tx,Das:2003mf}.
  
\section{Spontaneous symmetry breaking and restoration}
\subsection{NJL model on the light-front}
\label{sec:model}

The Nambu-Jona-Lasinio (NJL) originally suggested
in~\cite{Nambu:tp,Nambu:fr} has been reviewed in
Ref.~\cite{klevansky:1992} as a model of quantum chromo dynamics (QCD),
where also a generalization to finite temperature and finite chemical
potential has been discussed. Its generalization to the light-front including
a proper description of spontaneous symmetry breaking, which is not trivial,
has been done in Ref.~\cite{Bentz:1999gx}, which we use here. The
Lagrangian is given by
\begin{equation}
  \label{eqn:NJL}
  {\cal L}=\bar\psi (i\gc \partial -m_0)\psi 
+ G\left( (\bar\psi\psi)^2 + (\bar\psi i\gc_5 \tau\psi)^2\right). 
\end{equation}
In mean field approximation the gap equation is
\begin{equation}
  \label{eqn:gap}
  m=m_0 - 2G\langle \bar\psi\psi\rangle
= m_0 + 2 iG\gl\int\frac{d^4k}{(2\pi)^4}\;\mathrm{Tr}S_F(k),
\end{equation}
where $\gl=N_f N_c$ in Hartree and $\gl=N_f N_c+{\textstyle\frac{1}{2}}$ in
Hartree-Fock approximation, $N_c$ ($N_f$) is the number of colors (flavors).
For the isolated case $S_F(k)$ is the Feynman propagator.  Taking only the
lowest order in $1/N_c$ expansion of the 1-body or 2-body operators the light
front gap equation can be achieved by a $k^-$ integration, where in addition
$m_0\rightarrow \tilde m_0$ and $G\rightarrow \tilde G$ have to be
renormalized to accommodate the expansion. For details see~\cite{Bentz:1999gx}.
The propagator to be used in (\ref{eqn:gap}) is given in (\ref{eqn:lfmed}). The
gap equation becomes
\begin{equation}
  m(T,\mu)=\tilde m_0+2\tilde G\gl\int 
\frac{dk^+d^2k_\perp}{2k^+(2\pi)^3} \;4m(T,\mu)
(1-f^+(k^+,\vec k_\perp)-f^-(k^+,\vec k_\perp)).
  \label{eqn:gapmed}
\end{equation}
To regularize (\ref{eqn:gapmed}) we require $ k^-_\mathrm{on}+
k^+<2\Omega$.  As a consequence $k^+_1<k^+<k^+_2$ and
\begin{eqnarray}
\vec k^2_\perp &<& 2\Omega k^+- (k^+)^2 -m^2,
\label{eqn:regperp}
\\
k^+_{1,2}&=&\Omega\mp\sqrt{\Omega^2-m^2}.
\label{eqn:reglong}
\end{eqnarray}
For the isolated case this regularization is fully equivalent to the
Lepage-Brodsky one and the three-momentum cut-off. For the in medium case this
$\Omega$ regularization leads to analytically the same expressions as given
in~\cite{klevansky:1992} for the instantaneous case~\cite{Beyer:2003qb}.  The
calculation of the pion mass $m_\pi$, the pion decay constant $f_\pi$, and the
condensate value are also available on the light-front~\cite{Bentz:1999gx}.

\subsection{Results}
\label{sec:results}

 \begin{figure}[tbp]
\begin{minipage}[t]{0.49\textwidth}
   \begin{center}
     \epsfig{figure=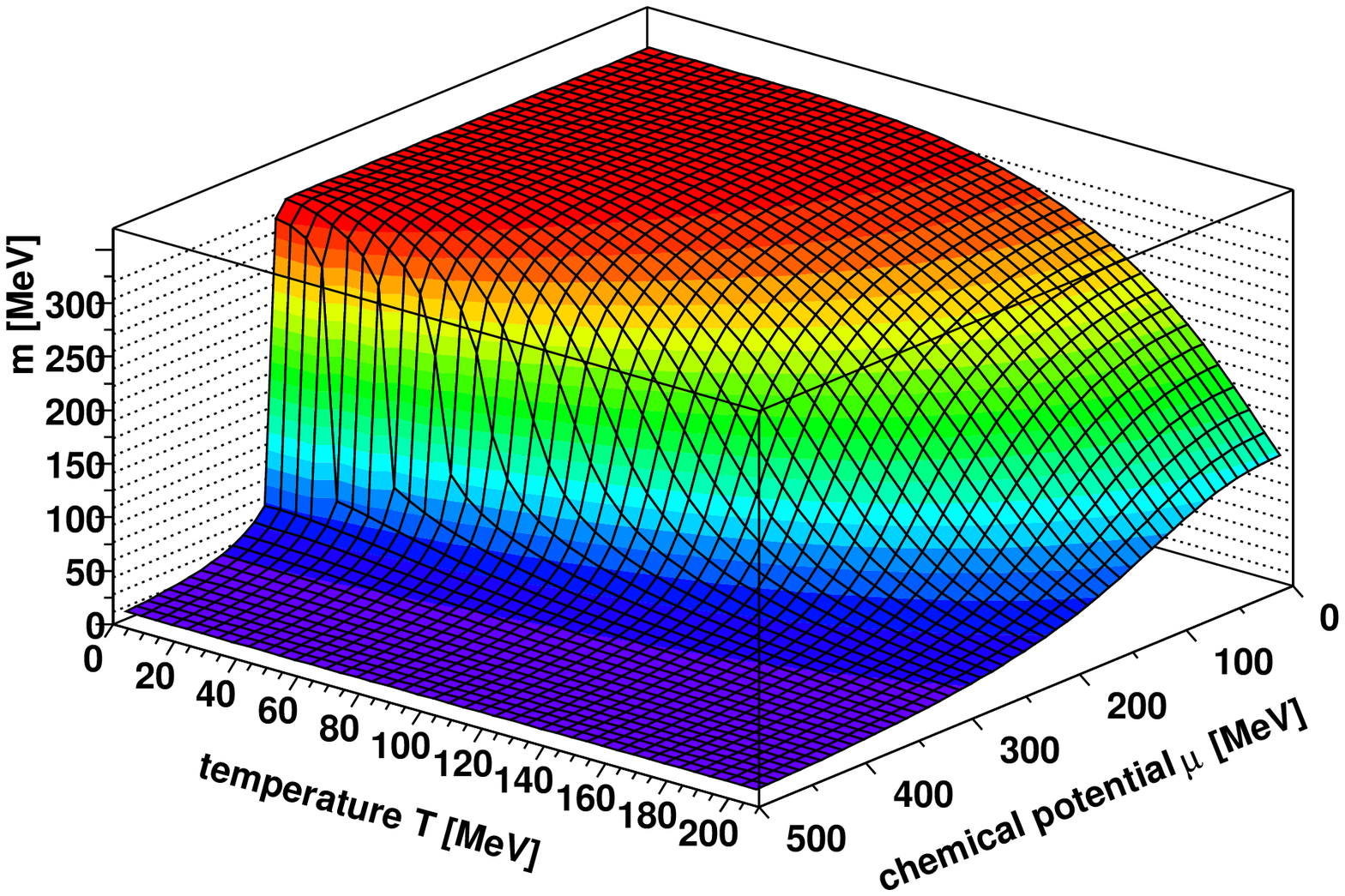, width=\textwidth}
     \caption{\label{fig:mass} 
       Effective quark mass as a function of temperature and chemical
       potential. The fall-off is related to the vanishing condensate
       $\langle \bar u u\rangle$, which shows the onset of chiral
       symmetry restoration.  Critical temperature at $\mu=0$ is
       $T_c\simeq 190$ MeV.}
   \end{center}
 \end{minipage}
\hfill
\begin{minipage}[t]{0.49\textwidth}
   \begin{center}
     \epsfig{figure=Tmu.eps, width=\textwidth}
     \caption{\label{fig:phase} 
       Chiral phase transition as defined in~\cite{asakawa:1989}. The lower
       part is the chiral broken phase, whereas the upper part reflects the
       restored phase.}
   \end{center}
 \end{minipage}
\end{figure}

The model parameters are adjusted to the isolated system. We use the Hartree
approximation, i.e. $\gl=N_cN_f=6$. Parameter values are chosen to reproduce
the pion mass $m_\pi=140$ MeV, the decay constant $f_\pi=93$ MeV, and to give
a constituent quark mass of $m=336$ MeV, i.e. $\tilde G =5.51\times
10^{-6}$~MeV, $\tilde m_0=5.67$~MeV, and $\Omega=714$~MeV.  The parameters are
reasonably close to the cases used in the review Ref.~\cite{klevansky:1992}.
The difference between the bare mass $\tilde m_0$ and the constituent mass is
due to the finite condensate, which is $\langle \bar uu\rangle^{1/3}=-247$
MeV.

In hot and dense quark matter the surrounding medium leads to a change of the
constituent quark mass due to the quasiparticle nature of the quark.  The
constituent mass as solution of (\ref{eqn:gapmed}) is plotted in
Fig.~\ref{fig:mass} as a function of temperature and chemical potential. The
fall-off is related to chiral symmetry restoration, which would be complete
for $m_0=0$. It is related to the QCD phase transition. For $T\lesssim 60$ MeV
the phase transition is first order, which is reflected by the steep change of
the constituent mass. To keep close contact with the 3M results we have chosen
for the $\Omega$ in-medium regulator mass
$\Omega^2(T,\mu)=\Lambda_\mathrm{3M}^2+m^2(T,\mu)$ with
$\Lambda_\mathrm{3M}=630$ MeV fixed for all $T$ and $\mu$.

We define the phase transition to occur at a temperature at which $m(T,\mu)$
is half of the isolated constituent quark mass~\cite{asakawa:1989}. The phase
diagram is shown in Fig.~\ref{fig:phase}. The line indicates the phase
boundary separating the hadronic phase from the quark gluon plasma phase.
Results presented in this section are based on an effective interaction in the
$q\bar q$ channel. They have to be supplemented by the medium dependence of
$f_\pi$ and $m_\pi$ that are currently underway.

\section{Few-particle correlations}
We are now interested in the $qq$ channel. Since we are going up to the
three-particle system, we presently approximate the spin structure.  To solve
the full three-fermion problem on the light front even for the isolated case
is quite a challenge. The spin structure is already rather complex, see
e.g.~\cite{Beyer:1998xy}.  Therefore the elementary spins are averaged
Tr$\gamma=0$ (in the medium) and hence, for the time being, we are only
dealing with bose type particles however subject to Fermi-Dirac statistics.
Our main focus here is to see how such a three-particle system is dynamically
influenced by a medium of finite temperature and density; ultimately, how
nucleons are formed in the hot and dense environment of a plasma of quarks and
gluons as the temperature and the density becomes smaller and how the relevant
degrees of freedom in the Fermi function change as the many-particle system
undergoes a change to hadronic degrees of freedom. To this end we need to
formulate suitable few-body equations that describe clusters of quarks in a
medium. In addition, because of the drastic mass change, see
Fig.~\ref{fig:mass}, these equations have to be relativistic ones.  The
equations derived here are based on a systematic quantum statistical framework
formulated on the light front using a cluster expansion for the Green
functions.  The formalism has been given elsewhere~\cite{Beyer:2001bc}.  We
repeat here the basics to make a connection to the previous sections.  The
light-front time ordered cluster Green function is defined by
\begin{equation}
i{\cal G}_{\ga\gb}(x-y)=
\theta(x^+-y^+)\;\langle A_\ga(x)\bar A_\gb(y)\rangle
\mp \theta(y^{+}-x^{+})\;\langle\bar A_\gb(y) A_\ga(x)\rangle.
\label{eqn:green}
\end{equation}
where all particles $A_\ga(x)\equiv
A_\ga(x^+,\uk{x})=\Psi_{\ga_1}(x^+,\uk{x_1})
\Psi_{\ga_2}(x^+,\uk{x_2})\Psi_{\ga_3}(x^+,\uk{x_3})\dots$ are all taken at
the same light front time $x^+$ and $\uk{x}=(x_+,\vec x_\perp)$. The upper
(lower) sign stands for fermion (boson) type clusters. Because of the global
light-cone time introduced, the dynamical equation for a cluster is equivalent
to a Dyson equation with a complicated mass operator that contains an
instantaneous part and a memory (or retardation) part. For the time being we
neglect the memory term. This is equivalent to a mean field approximation for
clusters leading to Faddeev-type three-body equations.

\begin{figure}[t]
  \begin{minipage}[t]{0.49\textwidth}
    \begin{center}
      \epsfig{figure=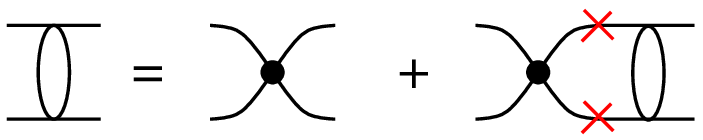,width=\textwidth}
      \caption{\label{fig:T2}
        Equation for the two-body $t$-matrix with zero
        range interaction. The crosses refer to the Pauli-blocking factor.}
    \end{center}
  \end{minipage}
  \hfill
  \begin{minipage}[t]{0.49\textwidth}
    \begin{center}
      \epsfig{figure=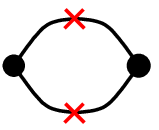,width=0.27\textwidth}
    \end{center}
    \caption{\label{fig:loop}
      Loop diagram corresponding to the kernel of the integral equation
      (\protect\ref{eqn:tau}). }
\vspace{3ex}
  \end{minipage}
\end{figure}
For a simple zero range interaction the $t$ matrix, Fig.~\ref{fig:T2},
separates and is given by the propagator $t(M_2)$, i.e.
\begin{equation}
t(M_2)=\left(i\gl^{-1} - B(M_2)\right)^{-1}.
\label{eqn:tau}
\end{equation}
The expression for $B(M_2)$ is represented by the loop diagram of
Fig.~\ref{fig:loop} and, in the rest system of the two-body system
$P^\mu=(M_2,M_2,\vec 0_\perp)$, given by
\begin{equation}
B(M_2)=-\frac{i}{(2\pi)^3} \int \frac{dx d^2k_\perp}{x(1-x)}
\frac{1-f(x,\vec k^2_{\perp})-f(1-x,\vec k_\perp^2)}{M_2^2-M_{20}^2},
\label{eqn:B}
\end{equation}
where $M_{20}^2=(\vec k_\perp^2+m^2)/x(1-x)$ and $f\equiv f^-$ given
in~(\ref{eqn:fermipm}) with $x=k^+/P^+_2$ where $\vk{P}_2=\vk{k}_1+\vk{k}_2$.
For a fermi system there are two important effects occurring due to the
blocking factors of (\ref{eqn:B}). One is the dissociation limit (Mott effect)
where $M_2(T_d,\mu_d)=2m(T_d,\mu_d)$. Above a certain temperature and density
no bound states can be formed. The second effect is related to the
appearance of a bose pole in the $t$ matrix. This happens for
$M_2(T_c,\mu_c)=2\mu_c$ and defines the critical temperature below which the system
becomes unstable and forms a new vacuum consisting of Cooper pairs
or a condensate. This is related to
superconductivity or superfluidity. In this case for $M_2^2\rightarrow
M^2_{02}$, we get
\begin{equation}
f(x,\vec k^2_{\perp})\Big|_{M^2_{02}\rightarrow M_2^2=4\mu^2}
=f(1-x,\vec k_\perp^2)\Big|_{M^2_{02}\rightarrow M_2^2=4\mu^2}=\frac{1}{2},
\end{equation}
i.e. both nominator and denominator of (\ref{eqn:B}) are zero.

The three-particle case is driven by the
Fadeev-type in medium equation 
\begin{equation}
\Gamma(y,\vec q_\perp) = \frac{i}{(2\pi)^3}\ t(M_2)
\int \frac{dxd^2k_\perp}{x(1-y-x)}
\frac{1-f(x,\vec k^2_\perp)
-f(1-x-y,(\vec k + \vec q)^2_\perp)}
{M^2_3 -M_{03}^2}\;\Gamma(x,\vec k_\perp),
\label{eqn:fad}
\end{equation}
where we have introduced vertex functions $\Gamma$ and $t(M_2)$ given before,
and an invariant cut-off $M_{30}^2<\gL^2$. Here the mass of the virtual
three-particle state (in the rest system $P^\mu=(M_3,M_3,\vec 0_\perp)$ is
\begin{equation}
M_{03}^2=\frac{\vec k^2_\perp+m^2}{x}
+\frac{\vec q^2_\perp+m^2}{y}
+\frac{(\vec k+\vec q)^2_\perp+m^2}{1-x-y},
\end{equation}
which is the sum of the on-shell minus-components of the three particles. The
nucleon scale is introduced by setting $M_3=938$ MeV. The isolated quark mass
used in these calculations is $m=386$ MeV. 

\begin{figure}[tbp]
\begin{minipage}[t]{0.49\textwidth}
   \begin{center}
     \epsfig{figure=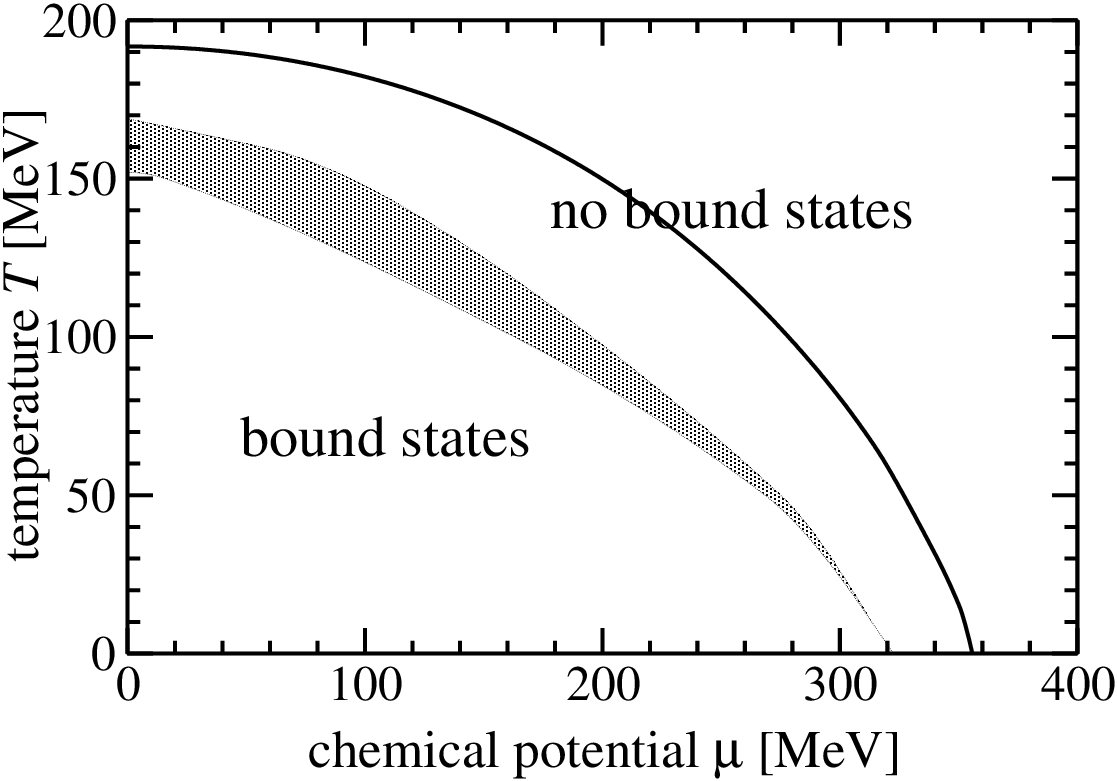, width=\textwidth}
     \caption{\label{fig:Mott} 
       Nucleon dissociation region (shaded area due to different
       cut-offs). Solid line chiral phase transition of Fig.~\ref{fig:phase}.}
   \end{center}
 \end{minipage}
\hfill
\begin{minipage}[t]{0.49\textwidth}
   \begin{center}
     \epsfig{figure=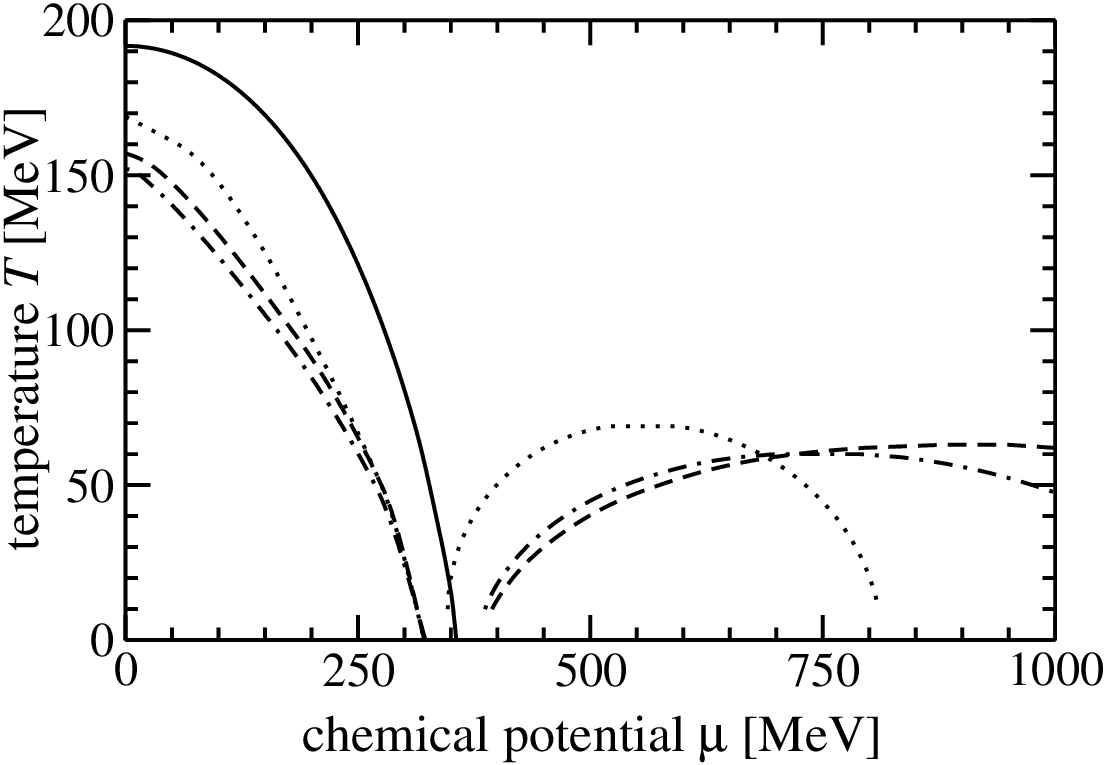, width=\textwidth}
     \caption{\label{fig:csc} 
       Phase diagram supplemented with critical temperature for color
       superconductivity. Different cut-offs: $\gL=4m$ (dot), $\gL=6m$ (dash),
       $\gL=8m$ (dash-dot).}
\vspace*{9ex}
   \end{center}
\end{minipage}
 \end{figure}

Fig.~\ref{fig:Mott} shows a shaded area that reflect the region where the
transition from baryons to quarks (or quark diquarks) occur. The area is
defined by use of different regularization masses. The chiral phase transition
given before is indicated by the solid line. Fig.~\ref{fig:csc} shows the
possible transition of quark matter to a superconducting phase.

\section{Conclusion and Outlook}
\label{sec:conclusion}
We have given a relativistic formulation of field theory at finite
temperatures and densities utilizing the light front form. The proper partition
operator (and the statistical operator) have been given for the grand
canonical ensemble. The special case of a canonical ensemble is given for
$\mu=0$. The resulting Fermi function depends on transverse and also on the
$k^+$ momentum components. The $k^+$ components emerge in a natural way in a
covariant approach. As an application we have revisited the NJL low energy
model of QCD. We reproduce the phenomenology of the NJL model, in particular
the gap-equation and the chiral phase transition.  We have further given
consistent relativistic three-quark equations valid in a dense medium of
finite temperature. We find that the dissociation transition and the critical
temperature for the color superconductivity agree qualitatively with results
expected from other sources. However, the latter results are by no means
final. We have shown that it is possible to write down meaningful consistent
equations to solve the relativistic in-medium problem on the light front. The
next steps would be to use the NJL model all through to give a consistent
picture for the $q\bar q$ and the $qq$ channel. Further insight into this just
emerging possibilities of treating relativistic many-particle systems on the
light front might be provided by other theories, like 1+1 QCD, the Yukawa
model, and finally real QCD.

\paragraph{Acknowledgment:}
I gratefully acknowledge the fruitful collaboration with S. Mattiello, T.
Frederico, and H.J. Weber, who have substantially contributed to the work I
had the pleasure to present in this talk. I also thank S. Brodsky for his
interest and discussion on this new approach. In particular, I would like to
thank the organizers of the workshop for a fruitful and well organized
meeting. Work supported by Deutsche Forschungsgemeinschaft, grant BE 1092/10.

\end{document}